\documentclass[prd,twocolumn,showpacs]{revtex4-1}
\usepackage[utf8x]{inputenc}
\usepackage{graphicx}
\usepackage{amsmath,amssymb} 
\usepackage[english]{babel}
\usepackage{rotating}
\usepackage{amsbsy}
\usepackage{textcomp}
\usepackage{psfrag}
\usepackage{multirow}
\usepackage{color} 
\usepackage{sidecap}
\usepackage{array}
\usepackage{xspace}

 \newcommand{\beq }{\begin{equation}}
\newcommand{\eeq}{ \end{equation}}
\newcommand{\beqa }{\begin{eqnarray}}
\newcommand{\eeqa }{\end{eqnarray}}
 \newcommand{\bwt }{\begin{widetext}}
 \newcommand{\ewt }{\end{widetext}}
 \newcommand{\bef}{\begin{figure}[h!]}
\newcommand{\eef}{\end{figure}}

\newcommand{\apjl}{\emph{Astrophys. J. Letters \xspace}}

\newcommand{\lif}{\texttt{LALInference}\xspace} 
\newcommand{\gws}{GWs\xspace}
\newcommand{\gw}{GW\xspace}
\newcommand{\lvc}{LIGO-Virgo Collaboration\xspace}

\newcommand{\EOS}{EOS\xspace}
\newcommand{\nn}{\newline}
\newcommand{\mc}{$\mathcal{M}$\xspace}
\newcommand{\sqdeg}{deg$^2$\xspace}
\newcommand{\msun}{$M_\odot$\xspace}

\newcommand{\mchirp}{$\mathcal{M}$\xspace}
\newcommand{\mratio}{$Q$\xspace}
\newcommand{\ra}{$\alpha$\xspace}
\newcommand{\dec}{$\delta$\xspace}
\newcommand{\pol}{$\psi$\xspace}
\newcommand{\phic}{$\phi_{c}$\xspace}
\newcommand{\timec}{$t_c$\xspace}
\newcommand{\dtwo}{Deg$^2$\xspace}
\newcommand{\PlotPath}{.}

\newcommand{\si}{$\sim$\xspace}
\newcommand{\sq}{squeezing\xspace}
\newcommand{\qsq}{quantum squeezing\xspace}
\newcommand{\LS}{``Lossy''\xspace}
\newcommand{\LL}{``Lossless''\xspace}
\newcommand{\BS}{``Baseline''\xspace}
\newcommand{\SQ}{``Squeezed''\xspace}

\newcommand{\adlv}{Advanced LIGO and Virgo\xspace}

\newcommand{\hoff}{$h(f)$\xspace}
\usepackage{aas_macros}
\newcommand{\vtheta}{$\vec\theta$\xspace}
\newcommand{\vd}{$\vec d$\xspace}
\newcommand{\post}{$p(\vec\theta | \vec d)$\xspace}

\begin{document}
\title{Effect of squeezing on parameter estimation of gravitational waves emitted by compact binary systems}
\author{Ryan Lynch}\email{ryan.lynch@ligo.org}
\author{Salvatore Vitale}
\author{Lisa Barsotti}
\author{Matthew Evans}
\affiliation{Massachusetts Institute of Technology, 185 Albany St, 02138 Cambridge USA}
\author{Sheila Dwyer}
\affiliation{LIGO Hanford Observatory, PO Box 159, Richland, WA 99352, USA}
\begin{abstract}

The LIGO gravitational wave (GW) detectors will begin collecting data in 2015, with Virgo following shortly after. These detectors are expected to reach design sensitivity before the end of the decade, and yield the first direct detection of GWs before then.  The use of squeezing has been proposed as a way to reduce the quantum noise without increasing the laser power, and has been successfully tested at one of the LIGO sites and at GEO in Germany. When used in Advanced LIGO without a filter cavity, the squeezer improves the performances of detectors above \si100 Hz, at the cost of a higher noise floor in the low frequency regime. Frequency-dependent squeezing, on the other hand, will lower the noise floor throughout the entire band. Squeezing technology will have a twofold impact: it will change the number of expected detections and it will impact the quality of parameter estimation for the detected signals. In this work we consider three different GW detector networks, each utilizing a different type of squeezer -- all corresponding to plausible implementations. Using LALInference, a powerful Monte Carlo parameter estimation algorithm, we study how each of these networks estimates the parameters of GW signals emitted by compact binary systems, and compare the results with a baseline advanced LIGO-Virgo network.
We find that, even in its simplest implementation, squeezing has a large positive impact: the sky error area of detected signals will shrink by \si 30\% on average, increasing the chances of finding an electromagnetic counterpart to the GW detection. Similarly, we find that the measurability of tidal deformability parameters for neutron stars in binaries increases by \si30\%, which could aid in determining the equation of state of neutron stars. 
The degradation in the measurement of the chirp mass, as a result of the higher low-frequency noise, is shown to be negligible when compared to systematic errors. Implementations of a quantum squeezer coupled with a filter cavity will yield a better overall network sensitivity. They will give less drastic improvements over the baseline network for events of fixed SNR but greater improvements for identical events.

\end{abstract}
\maketitle

\section{Introduction}\label{Sec.Intro}

The era of ground-based gravitational wave astronomy is about to begin. The Advanced LIGO~\cite{Harry:2010zz} detectors are expected to come online in 2015~\cite{2013arXiv1304.0670L}, whereas Advanced Virgo~\cite{AVirgo} should start taking data in 2016~\cite{2013arXiv1304.0670L}. 
Through a sequence of commissioning and observing periods, the advanced detectors should reach their design sensitivities over the next 3-4 years.
Two additional instruments, LIGO India~\cite{Indigo} and the Japanese Kagra~\cite{2012CQGra..29l4007S}, should join the global network of gravitational wave detectors before the end of the decade, further increasing its sensitivity.

Several astrophysical phenomena are known which should produce gravitational waves (\gws) measurable with ground-based detectors. The most promising sources are compact binary coalescences (CBCs) made of neutron stars and/or black holes. Once at design sensitivity, advanced detectors are expected to detect \si70 CBCs per year (although this rate has significant uncertainties~\cite{2010CQGra..27q3001A}).
Analysis of detected signals will broaden our understanding of compact objects and binary formation. For example, mass measurements can give insight into the mass distribution of neutron stars and black holes in binaries, and could reveal or dismiss the presence of a ``mass gap'' between the largest neutron stars and smallest black holes~\cite{2012ApJ...757...36K}.   Measurements of neutron star tidal deformability may help constrain the equation of state of matter in extreme conditions~\cite{2013PhRvL.111g1101D,2014PhRvD..89j3012W}. 
\gws will be used to measure the spin of black holes and neutron stars~\cite{PhysRevLett.112.251101,2014arXiv1403.0544O,2014arXiv1404.3180C}, which may help shed light into the evolutionary paths of binary systems and verify how efficiently common envelope evolution aligns spins with the system's orbital angular momentum.

While most of the effort is currently being put into preparing the advanced detectors for the first observing period (late 2015~\cite{2013arXiv1304.0670L}), research and development continue to improve this generation of ground-based detectors, and shape the next one~\cite{2010CQGra..27s4002P,2014arXiv1410.0612D}. Quantum noise will dominate throughout the detection band of \adlv, with thermal noise contributing significantly below 100Hz. 

Squeezing has been proposed as a mean of reducing the quantum noise of advanced detectors without having to increase the laser power~\cite{LIG11b,Bar13a,Grote2013}. In its simplest implementation (frequency-independent \sq), quantum \sq lowers the medium- and high-frequency noise floor of the detector at the expense of the low-frequency noise floor (see Fig.~\ref{Fig.NoiseCurves} top panel). Further developments will couple the squeezer with a filter cavity to control the squeezing in a frequency-dependent fashion, avoiding the low-frequency sensitivity degradation produced by frequency independent squeezing (see Fig.~\ref{Fig.NoiseCurves} bottom panel), with respect to the baseline noise spectral density (defined on Sec.~\ref{SubSec.NoiseModel})~\cite{Kim01a,Eva13a,Kwee14a,Che05a}.
Although squeezing was not part of the baseline configuration for the Advanced LIGO detectors, encouraging tests done so far (including at one of the LIGO sites~\cite{Bar13a}) suggest that at least the simpler squeezer without filter cavity may be mounted on the LIGO detectors already (as opposed to third-generation GW detectors).
It is thus interesting to determine if and to what extent quantum \sq can help gravitational wave astrophysics.

As mentioned above, frequency-independent squeezing reduces the noise at high frequencies while degrading the sensitivity at lower frequencies. It is important to stress that these two effects may somewhat balance each other when it comes to assessing the overall sensitivity of the detector (or network of detectors). It may thus be the case that the expected number of \gw detections and the range~\footnote{This is the volume and orientation averaged distance at which a (1.4,1.4)\msun binary neutron star would produces a single-interferometer signal-to-noise ratio (SNR) of 8, usually considered the threshold for detection.} of a network of interferometers do not strongly depend on whether squeezing is used or not. 
We will see that for the plausible noise spectral density we used, this is the case. As a result, the simplest implementation of quantum \sq will not lead to more frequent \gw detections.
However, the number of detections is not, obviously, the only figure of merit one should use to decide on the usefulness of \sq. By the time squeezers may realistically be implemented in ground based \gw detectors, it is very likely that one or several detections will have been made. On the other hand, \sq can change what we can learn about the astrophysical sources of detected \gws. 
\nn

The purpose of this study is to investigate how the implementation of squeezing in a LIGO-Virgo network affects  parameter estimation capabilities for CBC sources. We consider a baseline network consisting of LIGO and Virgo, and compare it to three hypothetical networks where the LIGO instruments contain a squeezer (a frequency-independent squeezer; a frequency-dependent squeezer with a lossy filter cavity; and a frequency-dependent squeezer with a lossless filter cavity)~\cite{Evans:2013bs,Kwee:2014cd}.

We simulate CBC signals emitted by binary neutron star (BNS) and binary black hole (BBH) events and detected by the networks above, and verify how the different shapes of the noise floors due to \sq affect the quality of reconstruction of some key astrophysical parameters, such as the mass and the sky position of the \gw source. While comparing events across network configurations, we keep the same SNRs so that the differences we see are only due to the noise floor of the detectors and not the loudness of the source. 
We find that a network implementing frequency-independent squeezing in LIGO improves sky localization precision by \si 30\% with respect to the baseline advanced detector network. The measurability of neutron star tidal deformability improves by a similar amount.  These improvements come with a negligible degradation of the network's overall sensitivity and of the measurability of chirp mass (which we find to be limited by systematic errors). 

When filter cavities are used, the detectors with squeezing are more-or-equally sensitive than the baseline detector throughout the entire frequency band. Unlike the simpler frequency-independent squeezing, there won't be a trade-off between high-frequency and low-frequency noise, and we thus find that the total network sensitivity increases. 

This study suggests that quantum squeezing, even in its simplest implementation without a filter cavity, can have a large impact on parameter estimation of CBC events, and increase the scientific payoff of LIGO and Virgo detections.

\section{Method}\label{Sec.Method}
\subsection{Noise Models}\label{SubSec.NoiseModel}

The noise power spectral density (PSD) of a \gw detector is defined as the autocorrelation of the noise~\cite{SathyaSchutzLRR}. Working in the Fourier domain, this is written as 

\beq S(f) = 2 E\left[n(f)n^*(f)\right] \label{Eq.PSD} \eeq 

where $n(f)$ is the Fourier transform of the noise, the star represents complex conjugation, and $E[\circ]$ denotes an ensemble average for a detector. 
The noise in \gw detectors is not fully stationary (for example, there are known variations between day and night) nor it is fully Gaussian. However we can safely assume that for stretches of data long enough to contain a \gw signal ($\lesssim$ minutes) the noise is stationary. 
Initial LIGO and Virgo were affected by non-Gaussian noise fluctuations (known as glitches), typically short ($\lesssim 1$ sec) and loud. Similar artifacts will almost certainly also affect Advanced LIGO and Virgo.
Work is ongoing to try to either remove glitches from the data, or to take them into account in the analysis~\cite{2013PhRvD..88h4044L}. We expect these efforts to be fully mature by the time the first few detections are made.
In what follows, we thus assume that the noise in LIGO and Virgo can be considered Gaussian and stationary.
Under those hypotheses, the noise PSD fully characterizes the frequency-dependent sensitivity of the detector at any given time. Finally, we assume that the noise is an additive process; if a \gw \hoff is present, the data will read $d(f)=n(f)+h(f)$.

As mentioned above, we consider four hypothetical networks of \gw detectors:
    \begin{itemize}
     \item ``Baseline'': Two Advanced LIGO with design sensitivity; Virgo with design sensitivity;
     \item ``Squeezed'': Two Advanced LIGO with frequency-independent squeezing; Virgo with design sensitivity;
     \item ``Lossy'' : Two Advanced LIGO with squeezing and lossy filter cavity; Virgo with design sensitivity;
     \item ``Lossless'' : Two Advanced LIGO with squeezing and lossless filter cavity; Virgo with design sensitivity;
    \end{itemize}

Even though Virgo is considering the possibility of adding a squeezer in the future, no PSD curves of potential squeezing implementation in Virgo were available at the time of our analysis. This is why in all four scenarios we gave Virgo its design sensitivity.

The PSDs we used are show in Fig.~\ref{Fig.NoiseCurves}. It is clear how in its simpler implementation (top panel, red dashed) \qsq will improve the sensitivity above \si 100~Hz, while doing worse at lower frequencies. 
The bottom panel shows how a filter cavity will either maintain the same sensitivity as the baseline in the tens of Hertz region (``Lossy'') or do better (``Lossless'') while reaching the same sensitivity as frequency-independent squeezing at high frequency.  The generation of these PSDs are explained in~\cite{Evans:2013bs}.

\bef
\includegraphics[width=0.5\textwidth]{\PlotPath/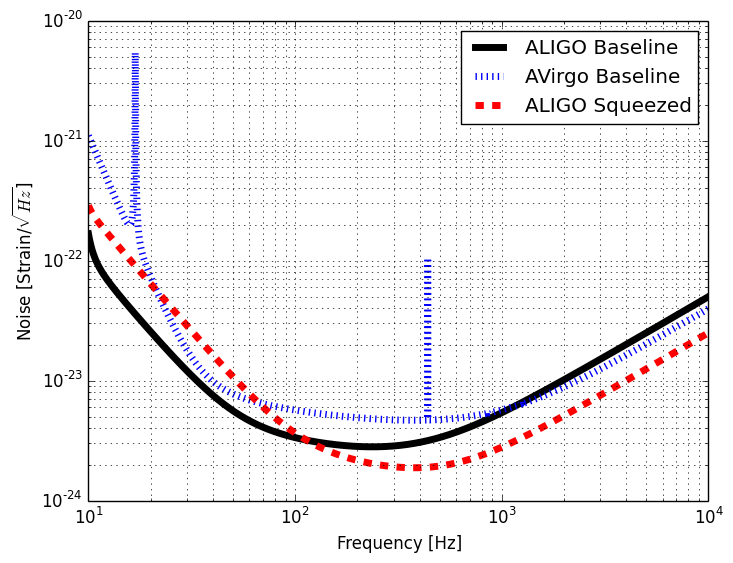}
\includegraphics[width=0.5\textwidth]{\PlotPath/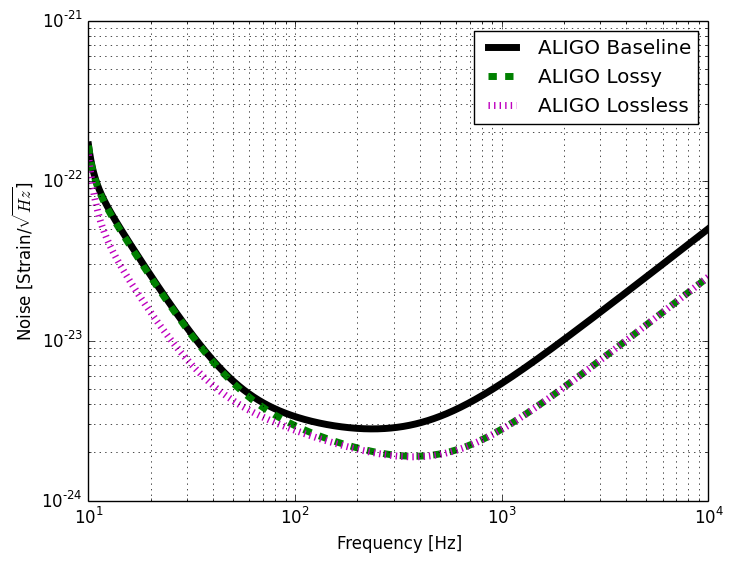}
\caption{(Top) The PSDs for the baseline LIGO (black solid) and Virgo (blue dotted) detectors and for a LIGO detector with frequency-independent squeezing (red dashed). (Bottom) The PSDs for a squeezed LIGO detector with lossy (green dashed) and lossless (magenta dotted) filter cavities; the design Advanced LIGO curve (solid line) is given for reference.}\label{Fig.NoiseCurves}
\eef

\subsection{Simulated \gw signals}\label{SubSec.Injections}

We considered 222 CBC sources, half of which consisted of two neutron stars (BNS), and the other half of two stellar-mass black holes (BBH).
We modeled BNS events using the frequency-domain TaylorF2 (TF2) waveform working at 3.5 Post-Newtonian (PN) phase order, while keeping a Newtonian amplitude order. TF2 waveforms can be explicitly written in the Fourier domain, see e.g.~\cite{2005PhRvD..71h4008A}. The waveforms were terminated at the innermost-stable circular orbit (ISCO) frequency~\cite{2005PhRvD..71h4008A}.
In Sec.~\ref{SubSubSec.Tides} we will report on the measurability of neutron star's tidal deformability. For those simulations, we included the known 5~PN and 6~PN tidal phase terms in the phase of TF2. Explicit expressions for these terms can be found in the appendix 5 of~\cite{2014PhRvD..89j3012W}.

To model \gw emitted by BBH, we used IMRPhenomB (IMRb) waveforms~\cite{2011PhRvL.106x1101A} with (anti-)aligned spins. 
IMRb are better suited for the larger masses of BBH, since the merger and ringdown phases~\cite{2011PhRvL.106x1101A} (which TF2 does not model) may be in a sensitive part of the detectors, and we will want to take them into account.
IMRb waveforms do have a phenomenological merger and ringdown, tuned against numerical simulations. Here again we worked at 3.5~PN phase order, while keeping a Newtonian amplitude order.

For both families of events, the sources were randomly distributed on the sky and given random orientations. Neutron star masses were generated uniformly in the realistic range $[1.4-2.3]$~\msun, whereas for black holes the range was $[5-25]$~\msun. 
Black holes were given random reduced spins~\footnote{Defined as $a\equiv|\vec{S}|/m^2$, where $\vec{S}$ and $m$ are the spin and the mass of the black hole} (along the direction of the orbital angular momentum) in the range $\pm[0.1-0.9]$, a negative sign indicating that spin and orbital angular momentum are anti-aligned.
Mass, position, orientation, and spin were kept fixed while analyzing events with different network configurations. 

The distances of the sources were uniform in volume, and thus represented an astrophysically realistic distribution.
We imposed a cut on the SNR of the sources, only analyzing signals with network SNR in the range $[12-40]$, with SNR 12 roughly corresponding to the threshold value for detection of a CBC signal.
The optimal network SNR, $\rho$, is defined as usual:

\beq 
\rho^2 = \sum\limits_{D \in \text{detectors}}4\int^{f_{high}}_{f_{low}} df \frac{|h^D(f)|^2}{S^D(f)} \label{Eq.SNR} 
\eeq 

where $h^D(f)$ and $S^D(f)$ are the waveform and one-sided PSD, respectively, at the D-th detector.  

We first generated the set of simulated events for the baseline network in the way just described. If one had to analyze the \emph{same} events with a different network (e.g. \LS) the resulting SNR could be different as a result of different network sensitivities. In this case it would be impossible to disentangle the effects of the different SNR from the effect of the \emph{shape} of the PSDs. Because of this, we modified the distances of simulated events in a network-dependent fashion so that the resulting SNR are the same across networks for corresponding events.
The differences we see in the parameter estimation capabilities are thus not due to a different SNR, but only on the way the SNR is distributed in the bandwidth of each detector. 

\subsubsection{Parameter estimation}

In order to extract the parameters of the simulated signals buried into interferometers's noise, we used \lif, the parameter estimation algorithm put in place by the \lvc~\cite{2014arXiv1409.7215V}.
Accurate parameter estimation of CBC signals can be dealt with using a Bayesian approach, which allows for any prior information about the problem on hand to be taken into account.

We are interested in the posterior distribution of the unknown source parameters \vtheta given the \gw data \vd: \post. \vd indicates the data of all inteferometers participating in the analysis; in our case, $\vec d \equiv \{d^H, d^L, d^V\}$

One can use Bayes' theorem to write the posterior distribution for \vtheta as:
    
    \beq\label{Eq.Posterior}
    p(\vec\theta | \vec d) \propto  p(\vec d| \vec\theta) p(\vec\theta)
    \eeq
    
 The first term on the RHS in Eq.~\ref{Eq.Posterior} is the likelihood of the data given the parameters, whereas the second one is the prior distribution of the parameters.
 Under our working hypothesis of stationary Gaussian noise, and taking into account that the noise in each detector is effectively independent from all others', the likelihood can be written as:
     
     \beq
     p(\vec d | \vec\theta) \propto \prod_{i=\{H,L,V\}}{e^{-\frac{1}{2}\langle d^i - h^i(\vec\theta) | d^i - h^i(\vec\theta)\rangle}}.
     \eeq
     
In the expression above, $d^i$ is the data of the i-th interferometer, $h^i(\vec\theta)$ is the waveform template (TF2 for BNS, IMRb for BBH) calculated with parameters $\vec\theta$, and the angular brackets represent a noise-weighted scalar product: $$\langle a|b \rangle\equiv  4\text{Re}\int{ df \frac{\tilde{a}^{*}(f)\tilde{b}(f)}{S(f)}}$$.

The prior distribution of \vtheta represents what is known of the CBC sources before the data is analyzed. We used isotropic priors on the sky position and orientation of the sources. The prior on the distance was uniform in volume, $p(D) \propto D^2$. The prior for all other parameters was uniform with prior bounds large enough to ensure the posterior distribution would not be cut~\footnote{One exception is the mass ratio. Since by convention \lif assumes $m_1>m_2$ the prior for the mass ratio $Q=m_1/m_2$ has a natural upper bound at 1.0.}.

\lif uses Monte Carlo methods to explore the multi-dimensional parameter space in an efficient and reliable way and to estimate the full multi-dimensional posterior distribution of \vtheta, from which all the interesting 1-D distributions can be obtained via marginalization. 
We used the nested sampling~\cite{2004AIPC..735..395S,2010PhRvD..81f2003V} flavor of \lif, in which one first calculates the Bayesian evidence for the signal model~\cite{2010PhRvD..81f2003V} and obtains the posterior distribution for \vtheta as a by-product.
We point the interested readers to~\cite{2014arXiv1409.7215V,2010PhRvD..81f2003V} for more details on the nested sampling technique and its implementation for \gw parameter estimation.

\section{Results}\label{Sec.Results}

\subsection{BNS}\label{SubSec.BNSresults}
In this section we report the results for BNS sources. 
We will first compare the \BS and \SQ networks, deferring to Sec.~\ref{SubSubSec.LossyLossless_SSNR} the analysis of the \LS and \LL networks.

In sec.\ref{SubSubSec.BNS_base_v_sqz} we will consider the simplest (and least computationally expensive) case of BNS systems without tidal deformability. In this case the WF depends on 9 unknown parameters. They are:
    
\begin{itemize}
 \item Chirp mass \mchirp and mass ratio \mratio. They are defined in terms of the two component masses as \mchirp$\equiv\left[\frac{m_1^3 m_2^3}{m_1+m_2}\right]^\frac{1}{5}$ and $q\equiv m_1/m_2$;
 \item Coalescence time \timec and phase \phic. These are the time of the coalescence and the phase of the WF at that time;
 \item Polarization \pol, defined in e.g.~\cite{1989JBAA...99..196H};
 \item Luminosity distance $D_L$;
 \item Right ascension \ra and declination \dec;
 \item Orbital inclination $\iota$, i.e. the angle between the orbital angular momentum and the line of sight.
\end{itemize}

Tidal parameters will be taken into account for a subset of events in Sec.~\ref{SubSubSec.Tides}, which adds 2 additional unknown parameters to the problem (see ~\cite{2014PhRvD..89j3012W}).

\subsubsection{Baseline vs. Squeezed}\label{SubSubSec.BNS_base_v_sqz}

As previously mentioned, we kept all source parameters, except for the distance, fixed while simulating events analyzed with different networks.
The first interesting question is by how much one needs to modify the distance of BNS events to find the same SNR in the different network configurations. This is shown in Fig.~\ref{Fig.DistancesBNS}.

It is clearly visible that the distributions of distances for the \BS and \SQ networks are nearly identical, and we calculate the average difference in the distances to be \si1\%. This implies that, from a detection point of view, there is no advantage in using the \SQ network.
Given the significantly different shape of PSDs for the baseline and squeezed LIGO (Fig.~\ref{Fig.NoiseCurves}, top), the fact that the two networks have the same sensitivity implies that the SNR lost by the \SQ configuration at low frequencies is nearly identically compensated for at high frequencies. This is confirmed in Table~\ref{Tab.BNS_SNR}, where we  show how the SNR is (on average) distributed among the inteferometers, and how it is distributed in four (arbitrarily chosen) frequency bins.

\begin{table*}[ht]
\centering
\scriptsize
\caption{The average SNR$^2$ distribution for BNS events, used to characterize the distribution of network sensitivity.  Each percentage is the ratio of the the squared SNR in each detector/bin to the square of the total network SNR, averaged over all events.  The first three columns describe the fraction of average squared SNR in the LIGO Hanford (H), LIGO Livingston (L), and Virgo (V) detectors, respectively.  The last four columns describe the fraction of squared network SNR in four frequency bins.}\label{Tab.BNS_SNR}
\begin{tabular}[c]{c||c|c|c||c|c|c|c}
PSD & H (\%) &  L (\%) & V (\%) & 30-60 Hz (\%) & 60-200 Hz (\%) & 200-512 Hz (\%) & 512~Hz-ISCO (\%)\\
\hline\hline
Baseline & 39.3 & 40.2 & 20.5 & 27.9 & 55.4 & 14.5 & 2.2\\
Squeezed & 39.1 & 40.0 & 20.9 & 12.5 & 52.3 & 29.1 & 6.1\\
Lossy & 41.5 & 42.3 & 16.2 & 21.9 & 52.7 & 21.0 & 4.4\\
Lossless & 42.7 & 43.5 & 13.8 & 29.2 & 49.5 & 17.7 & 3.6\\
\end{tabular}
\end{table*}

\begin{table*}[htb]
\scriptsize
\caption{The average parameter estimation capabilities for equal-network-SNR BNS events.  The first two columns give the average relative errors for the mass parameters \mc and Q. The next three columns give the average timing errors for each instrument, in milliseconds.  The last two columns give the average areas of the measured 90\% and 67\% confidence regions of source location for each detector.  All quantities are averaged over the entire ensemble of BNS events.}\label{Tab.BNS_pe}
\begin{tabular}[c]{c||c|c||c|c|c||c|c}
PSD & $\Gamma_\mathcal{M}$ (\%) & $\Gamma_Q$ (\%) & $\sigma_{t_H}$ (ms) & $\sigma_{t_L}$ (ms) & $\sigma_{t_V}$ (ms) & 90\% Conf (\sqdeg) & 67\% Conf (\sqdeg)\\
\hline\hline
Baseline & $1.4\times10^{-2}$ & 7.6 & 0.26 & 0.31 & 1.4 & 17 & 8.7\\
Squeezed & $1.6\times10^{-2}$ & 7.3 & 0.18 & 0.21 & 1.3 & 12 & 6.0\\
Lossy & $1.3\times10^{-2}$ & 7.1 & 0.19 & 0.23 & 2.2 & 17 & 8.2\\
Lossless & $1.2\times10^{-2}$ & 7.0 & 0.21 & 0.25 & 2.7 & 20 & 10\\
\end{tabular}
\end{table*}

\bef
\includegraphics[width=0.5\textwidth]{\PlotPath/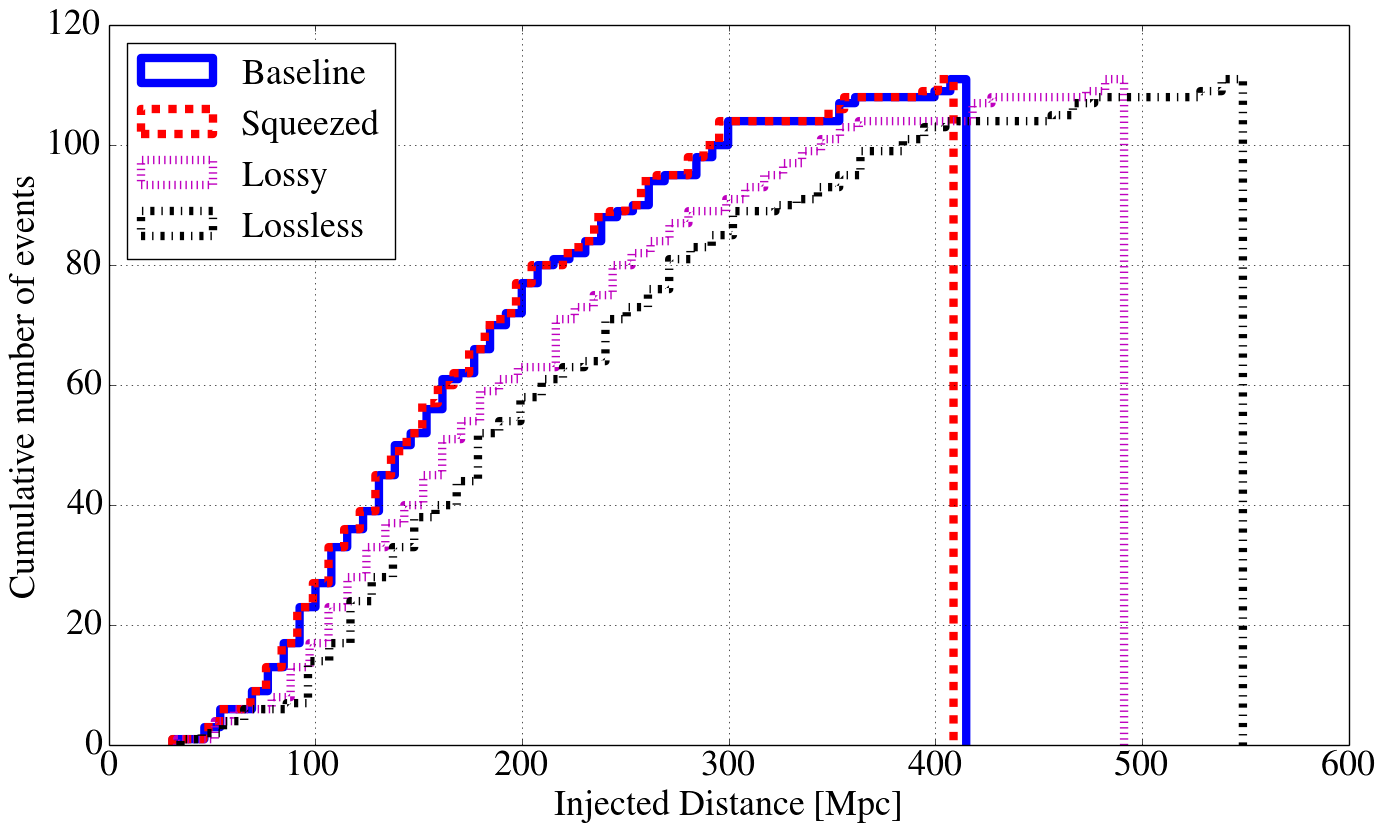}
\caption{A cumulative histogram of the injected distance for the BNS events for each network (where the SNR distribution is identical for all networks). Note that the distances of the squeezed events are very similar to those of the baseline events, which implies that the squeezed and baseline networks have very similar network sensitivities.  On the other hand, the distances of events for the lossy and lossless networks are further than for the baseline network, which implies that these two networks have greater network sensitivity (with the lossless network being most sensitive).}\label{Fig.DistancesBNS}
\eef

We see that, on average, each of the two LIGO instruments are responsible for \si40\% of the total squared SNR, in both the \BS and \SQ networks.
We also see how the relative importance of the low-frequency (30-60~Hz) and high-frequency (200Hz-ISCO) bins is flipped in the two configurations.  This is obviously expected, and quantifies the differences between the two PSDs we have qualitatively described above.

We can now consider parameter estimation. Before reporting our results, we can make a few general considerations that will help in predicting and understanding them.
Due to the higher noise floor at low frequencies, we may expect the \SQ network to do worse than the \BS for those parameters which enter the WF's phase at low Post-Newtonian orders.
Roughly speaking, this happens because the k-th PN orders gets multiplied by $f^\frac{k-5}{3}$~\cite{2005PhRvD..71h4008A}, which is larger at low frequencies when $k<5$ (i.e. for PN orders below the 2.5PN). Low PN terms should thus be more sensitive to the behavior of the instrument at low frequencies. For example, the chirp mass \mc already enters the WF at Newtonian order (i.e. 0~PN), while the mass ratio enters at 1~PN, and its measurement is significantly helped by higher PN terms (\cite{2005PhRvD..71h4008A}, Table II). We thus expect errors for the mass ratio to be unchanged or better when considering the \SQ network, whereas the measurability of \mc should degrade.

The opposite is true for the sky localization capabilities of LIGO-Virgo. Since GW detectors are approximately omnidirectional antennae, most information about the position of the sources comes from time triangulation~\cite{2014arXiv1404.5623S}.
Furthermore, the precision of the measurement of arrival time for CBC is more sensitive to the high-frequency part of the spectrum (see e.g. ~\cite{Fairhurst2009}). We thus expect the \SQ network to do better than the \BS in pinning down the position of BNS sources.

In Table~\ref{Tab.BNS_pe} we report the results of the \lif analysis. Specific to sky localization, we consider the 1-$\sigma$ errors in the measurement of the WF's arrival time and the 67\% and 90\% confidence interval for sky localization.
We can see how the increased high-frequency sensitivity of the LIGO detectors in the \SQ network leads to a significant improvement in the timing and sky localization measurements.  On average, the 90\% confidence level sky localization areas decreases from 17 \dtwo to 12 \dtwo while transitioning from the \BS to the \SQ network. That is a \si 30\% improvement. We notice how this matches with the improvement in the LIGO timing errors.  Given that BNS are expected to also be luminous in the electromagnetic spectrum~\cite{2010MNRAS.406.2650M,2012ApJ...746...48M}, a more precise sky localization will increase the chances of a joint EM-GW detection, which could boost the scientific payoff for both fields (see e.g.~\cite{2012A&A...539A.124L,2012A&A...541A.155A,2014ApJS..211....7A}).

Finally, we note that the timing errors in Virgo get slightly better. This is because the events in the \SQ scenario are slightly closer than in the \BS. The measured SNRs in Virgo are thus higher than in the \BS even though the network SNR is the same, leading to smaller errors. We will see later that the contrary happens for the \LS and \LL networks. 

We can now focus on the mass parameters. We report in Table~\ref{Tab.BNS_pe} the average 1-$\sigma$ error (in percent, relative to the true value) for \mc and Q. The trends confirm our qualitative guess: the chirp mass estimation gets less precise, while the contrary happens to Q. Even though these variations may look significant, in reality they are smaller than systematic errors due to WF uncertainties. 
It must, in fact, always be taken into account that the waveforms used for the analysis are only an approximate representation of what nature will produce. For example, it has been shown in~\cite{2013PhRvD..88f2001A} how the differences in the posterior distribution of \mc one obtains using different WF families are comparable to the 1-$\sigma$ errors of each posterior. To be more precise, typical systematic WF differences for BNS in ~\cite{2013PhRvD..88f2001A} are \si$0.1\%$, i.e. a factor of \si10 larger than the differences we find, due to quantum \sq. We will come back to this point later while discussing the BBH results.

In summary, for the simulated set of BNS signals considered in this section, the \SQ network achieves a significant improvement in sky localization capabilities, as compared to the baseline network, without losing any meaningful mass estimation capabilities and maintaining the same network sensitivity.

\subsubsection{Tidal Parameters}\label{SubSubSec.Tides}

We have seen that the numerical results we found in the previous section confirmed the trends we would have expected from simple qualitative considerations. In particular, we saw that parameters entering the PN series at higher orders benefited more from using the \SQ network. We may thus expect that the tidal deformability parameters~\cite{2008PhRvD..77b1502F}, which are formally 5~PN and 6~PN, would be measured significantly better by the \SQ network. Precise measurement of neutron stars' tidal deformability  has been shown to be possible with GW data alone~\cite{2014PhRvD..89j3012W,2013PhRvL.111g1101D}, and could be pivotal for a better understanding of the behavior of nuclear matter in extreme conditions.

To verify whether the measurability of tidal parameter would benefit from squeezing, we have compared the performances of the \BS and \SQ networks for a set of 13 BNS systems. We chose the simulated BNS events to have the same masses, orientations, equation of state, and SNRs as those used by Wade et al. in~\cite{2014PhRvD..89j3012W}. Similarly to~\cite{2014PhRvD..89j3012W} we chose to run the analysis with a zero-noise realization~\footnote{This means that, while the noise PSD is non-zero, the actual noise if zero in each time bin. Under the assumption of Gaussian noise, this can be seen as the average result one would get analyzing the same event under different noise realizations.}.
The parametrization of the tidal phase terms can be found in~\cite{2014PhRvD..89j3012W}. Here we will just mention that 2 new (unknown) parameters are necessary to describe the effects of tidal deformability ($\tilde{\Lambda}$ and $\delta\tilde{\Lambda}$), bringing the total number of unknown parameters to 11.  $\tilde{\Lambda}$ is the tidal deformability parameter.  It is a dimensionless reparametrization of each of the neutron stars' individual tidal deformabilities~\cite{2014PhRvD..89j3012W}.

\begin{table*}[htb]
\centering
\scriptsize
\caption{A comparison of the tidal deformability $\tilde{\Lambda}$ measurement capabilities for the baseline and squeezed networks.  The first three columns give the neutron star component masses and the value of $\tilde{\Lambda}$ used for each run, mimicking those presented in~\cite{2014PhRvD..89j3012W}.  The remaining columns compare the measured relative error $\Gamma_{\tilde{\Lambda}}$ for the baseline and squeezed networks for events of network SNR of 20 and 30.}\label{Tab.BNS_tides}
\begin{tabular}[c]{c|c|c||c|c||c|c||c|c}
\hline
\multicolumn{3}{c}{}& \multicolumn{2}{c}{$\rho=20$} & \multicolumn{2}{c}{$\rho=30$} \\
\hline
$m_1$ (\msun) & $m_2$ (\msun) & $\tilde{\Lambda}$ & $\Gamma_{\tilde{\Lambda}_{base}}$ (\%) &  $\Gamma_{\tilde{\Lambda}_{sqz}}$ (\%) & $\Gamma_{\tilde{\Lambda}_{base}}$ (\%) & $\Gamma_{\tilde{\Lambda}_{sqz}}$ (\%) \\
\hline\hline
1.20 & 1.20 & 1135.630 & 22 & 14 & 10 & 7 \\
1.35 & 1.20 & 820.610 & 30 & 17 & 16 & 10 \\
1.35 & 1.35 & 590.944 & 28 & 22 & 16 & 13 \\
1.50 & 1.35 & 435.585 & 35 & 28 & 22 & 18 \\
1.65 & 1.35 & 328.177 & 53 & 39 & 36 & 28 \\
1.50 & 1.50 & 318.786 & 42 & 31 & 29 & 21 \\
1.80 & 1.35 & 252.398 & 78 & 56 & 48 & 39 \\
1.95 & 1.35 & 197.899 & 110 & 70 & 60 & 46 \\
1.65 & 1.65 & 175.963 & 54 & 44 & 40 & 33 \\
2.10 & 1.35 & 157.974 & 140 & 79 & 69 & 52 \\
1.80 & 1.80 & 98.191 & 72 & 64 & 52 & 47 \\
1.95 & 1.95 & 54.670 & 120 & 85 & 80 & 63 \\
2.10 & 2.10 & 29.844 & 190 & 140 & 120 & 99 \\
\end{tabular}
\end{table*}

The results of the measurements of tidal parameters are given in Table~\ref{Tab.BNS_tides}. First, we notice how the results for the \BS network are consistent with those of~\cite{2014PhRvD..89j3012W}. They are not identical because we let the position of our sources vary over the sky (though the sky positions are identical for corresponding \BS and \SQ network events).
Similarly to~\cite{2014PhRvD..89j3012W}, none of our simulations were able to measure or constrain $\delta\tilde{\Lambda}$, both for the \BS and \SQ networks.

\bef
\includegraphics[width=0.5\textwidth]{\PlotPath/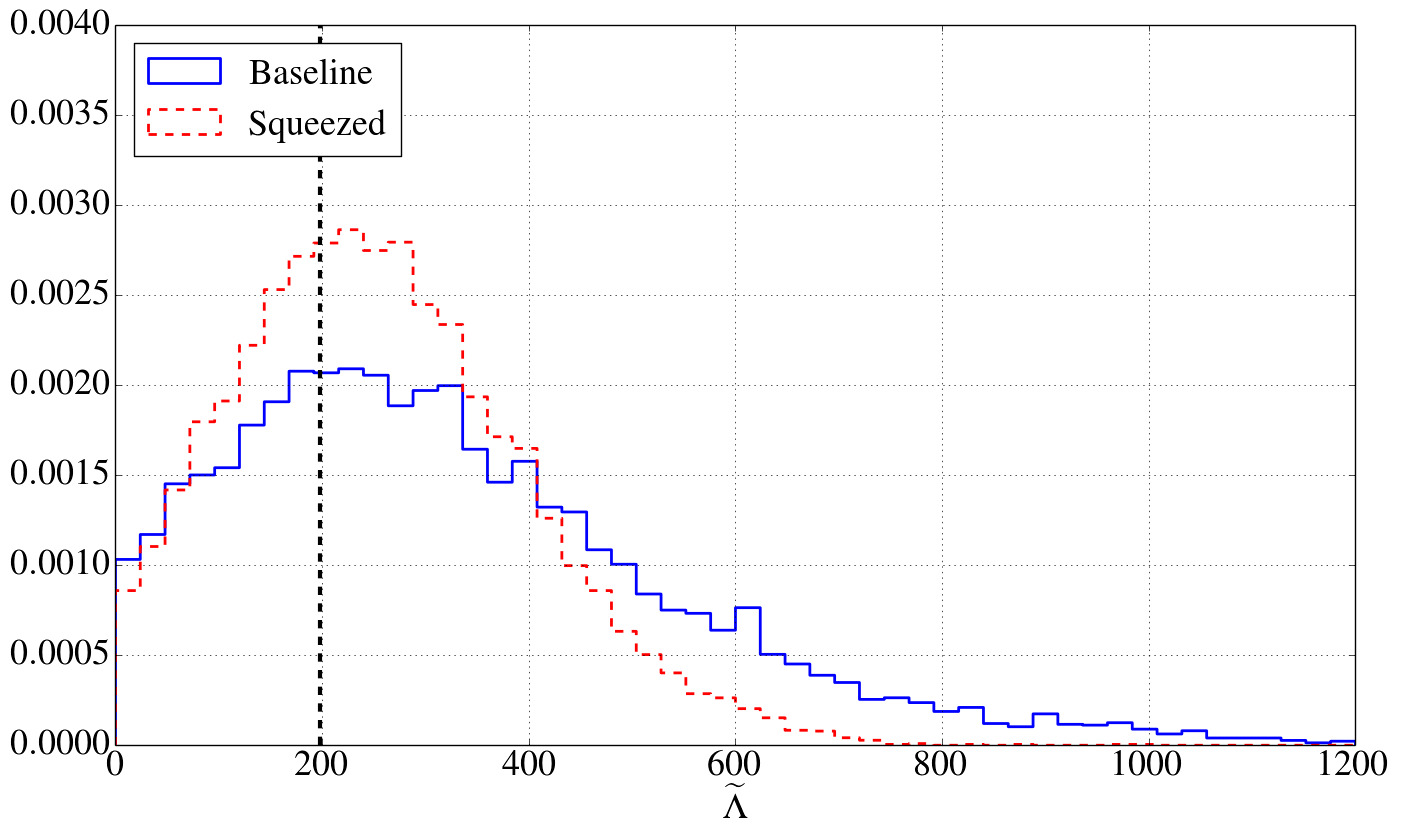}
\includegraphics[width=0.5\textwidth]{\PlotPath/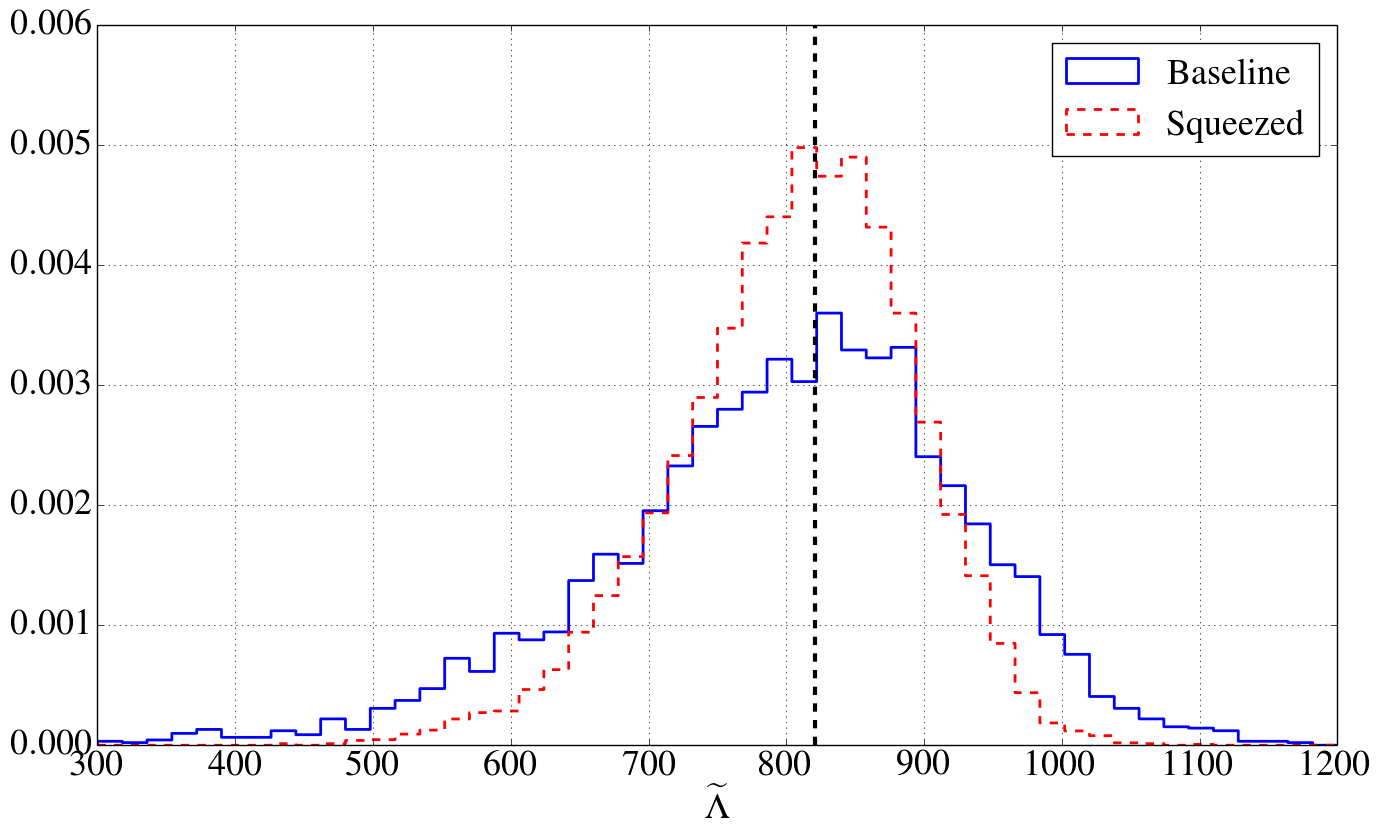}
\caption{Plots of the posterior density distributions of $\tilde{\Lambda}$ as recovered by the \BS (blue, continuous) and \SQ (red, dashed) networks, for the BNS system with $\tilde{\Lambda}= 197.899$ and SNR 20 (top) and the system with $\tilde{\Lambda}= 820.610$ and SNR 30 (bottom).}\label{Fig.TidalPost}
\eef

We see that the \SQ network performs better than the baseline network in measuring $\tilde{\Lambda}$, with average improvements of \si30\% for SNR 20 systems and \si20\% for SNR 30 systems.  Fig.~\ref{Fig.TidalPost} shows examples of the posterior distribution for individual events. These are improvements which could prove to be significant when placing EOS constraints.  However, it should be noted that because the phase of gravitational waveforms currently used for match-filtering is only known up to 3.5PN, whereas tidal parameters enter the GW phase at 5PN and 6PN, a bias may be introduced by the missing PN terms on the measured tidal parameters~\cite{2014PhRvD..89b1303Y}. Currently, this systematic waveform uncertainty is estimated at ~50\%~\cite{2014PhRvD..89j3012W}, which will be the dominant uncertainty in the measurement of $\tilde{\Lambda}$, even at high SNR. Work is ongoing to calculate the missing high-PN orders and reduce systematic errors.
Irrespective of eventual unresolved bias, it seems safe to assume that the \SQ network would yield a smaller statistical error than the \BS network for tidal measurements. 

\subsubsection{\LS and \LL}\label{SubSubSec.LossyLossless_SSNR}

We now present the results for the \LS and \LL networks. The bottom panel of Fig.~\ref{Fig.NoiseCurves} shows that the noise will be less-than-or-equal-to the \BS PSD across the entire bandwidth. Table~\ref{Tab.BNS_SNR} reveals that for the \LS network, an average of 75\% of the network SNR is located below 200 Hz. For the \LL network this number is 79\%. Comparing these numbers to those of the \BS network (83\%) and \SQ network (65\%), we see that both the \LS and \LL networks are more sensitive at high frequencies and less sensitive at low frequencies than the baseline network for equal SNR events, but these differences in sensitivity are not as large as for the \SQ network. 

We see from Table~\ref{Tab.BNS_pe} that the sky localization capability of the \LS network is similar to that of the \BS network, with average (90\%, 67\%) confidence level areas of sky location being (17, 8.2) \sqdeg. On the other hand, the \LL network performs worse than the \BS network, with average (90\%, 67\%) confidence level areas of sky location being (20, 10) \sqdeg. This can be explained by taking into account how the SNR is distributed across the frequencies and detectors.

We first notice that, since the two LIGO observatories are more sensitive in the \LS and \LL networks than in the \SQ and \BS networks and Virgo keeps the same PSD, Virgo's contribution to the network SNR is smaller. This is shown in Table~\ref{Tab.BNS_SNR}: Virgo's average percent contribution to the squared network SNR decreases from \si21\% for the \BS and \SQ networks to \si 15\% for the \LS and \LL. 
Since time triangulation yields important information about the sky position of the source, having a more uneven distribution of the SNR across the detectors hurts sky localization. 

At the same time, as seen above, the detector's high-frequency sensitivity also impacts sky localization. Thus, the increased high-frequency sensitivity of the LIGO detector in the \LS and \LL networks helps compensate for the increased imbalance of sensitivity among the detectors.  In summary, the better timing capabilities of the LIGO detectors (high-frequency sensitivity gains) is nullified by the worse timing capability of the Virgo detector (network sensitivity imbalance) to make the resulting sky areas of the \LS network comparable to the \BS network and the resulting sky areas of the \LL network worse than the \BS network.

From Table~\ref{Tab.BNS_pe} we see that the errors on the estimation of the mass parameters for the \LS and \LL networks are smaller than for the \BS network. 
It is worth stressing again that, especially for the chirp mass, we are already at the point where the uncertainty will be dominated by waveform systematics. Thus, the improvements to the mass parameters we found here will not be significant until better waveforms become available. 

What we described in this section may seem counter intuitive: we considered better LIGO instruments and found out that the parameter estimation precision was comparable (mass parameters) or worse (sky localization).
In reality, we must keep in mind that the \LS and \LL networks will have a larger horizon distance than the \BS network. This implies that, even though their performances may seem similar to the \BS for the \emph{average} BNS event, they will detect many more events at any given SNR (since \LS and \LL will be probing a larger volume of the universe).
The larger number of detections will have a significant impact in all those studies which rely on several tens of detected GWs to be successful (e.g. tests of General Relativity~\cite{2012PhRvD..85h2003L,2014PhRvD..89h2001A} and the equation of state of neutron stars~\cite{2013PhRvL.111g1101D}). For the same reasons, the probability of having a large SNR event is bigger for the \LL and \LS networks than for the \BS network.  In appendix \ref{App.SameD}, we give an example of what parameter estimation would look like for the \emph{same event} (i.e. keeping all parameters to be the same, including distance) when using the \LS and \LL networks.

\subsection{BBH Results}\label{SubSec.BBHresults}

\subsubsection{\BS and \SQ}\label{SubSubSec.BBH_base_v_sqz}

In our analysis of BNS systems we have seen how the distance range of the \SQ and \BS networks was the same. This holds because the better sensitivity at high frequency of the \SQ network was compensated for by the worsening of sensitivity in the tens of Hertz region. The situation could be different for more massive systems, since the maximum frequency of a CBC system in the LIGO-Virgo band is inversely proportional to the system's total mass~\cite{2005PhRvD..71h4008A}\footnote{As an example, while ISCO for a (1.4,1.4)\msun BNS happens at \si1570~Hz, the ringdown frequency for a (10,10)\msun BBH is only \si800~Hz (and the merger starts at \si350~Hz).}.
For massive enough systems it may be the case that the improvement at high frequencies is not fully taken advantage of since the waveforms end at lower frequencies than their BNS counterparts.

We find that, in order to achieve the same SNR, the BBH events must be typically placed closer for the \SQ network (by \si3\% on average), which implies that the \SQ network is only slightly less sensitive than the \BS to BBH over the mass range we considered (we don't show a plot since it would look qualitatively similar to Fig.~\ref{Fig.DistancesBNS}).

\begin{table*}[ht]
\centering
\scriptsize
\caption{The average SNR$^2$ distribution for equal-network-SNR BBH events, used to characterize the distribution of network sensitivity.  Each percentage is the ratio of the square of the SNR in each detector/bin to the square of the total network SNR, averaged over all events.  The first three columns describe the fraction of average SNR$^2$ in the Hanford, (H), Livingston (L), and Virgo detectors, respectively.  The last four columns describe the fraction of network SNR$^2$ in each frequency range.}\label{Tab.BBH_SNR}
\begin{tabular}[c]{c||c|c|c||c|c|c|c}
PSD & H (\%) &  L (\%) & V (\%) & 30-60 Hz (\%) & 60-200 Hz (\%) & 200-512 Hz (\%) & 512-1024 Hz (\%)\\
\hline\hline
Baseline & 39.3 & 40.1 & 20.6 & 30.4 & 54.3 & 14.0 & 1.3\\
Squeezed & 38.7 & 39.5 & 21.8 & 14.2 & 52.7 & 29.5 & 3.6\\
\end{tabular}
\end{table*}

\begin{table*}[ht]
\centering
\scriptsize
\caption{The average parameter estimation capabilities for equal-network-SNR BBH events.  The first two columns give the average relative errors for the mass parameters \mc and Q. The next three columns give the average timing errors for each instrument, in milliseconds.  The last two columns give the average areas of the measured 90\% and 67\% confidence regions of source location for each detector.  All quantities are averaged over the entire ensemble of BBH events.}\label{Tab.BBH_pe}
\begin{tabular}[c]{c||c|c||c|c|c||c|c}
PSD & $\Gamma_\mathcal{M}$ (\%) & $\Gamma_Q$ (\%) & $\sigma_{t_H}$ (ms) & $\sigma_{t_L}$ (ms) & $\sigma_{t_V}$ (ms) & 90\% Conf (\sqdeg) & 67\% Conf (\sqdeg)\\
\hline\hline
Baseline & 0.95 & 30 & 0.74 & 0.76 & 2.3 & 21 & 11\\
Squeezed & 1.2 & 25 & 0.61 & 0.64 & 2.0 & 15 & 7.3\\
\end{tabular}
\end{table*}

Table~\ref{Tab.BBH_SNR} shows how the SNR squared is distributed in 4 different frequency bins. A comparison with Table~\ref{Tab.BNS_SNR} makes it clear that the frequency-dependent sensitivity distribution is very similar to that of the BNS simulations, with only a slight shift of network sensitivities towards low frequencies. Thus, although solar-mass BBH signals are shorter than BNS signals, we do expect to find parameter estimation trends similar to what is seen for BNS signals.

In Table~\ref{Tab.BBH_pe} we report the average errors for the BBH mass parameters. The errors on the estimation of the chirp mass are larger than those for BNS (for corresponding networks). This is clearly to be expected, since BBH signals are shorter, and thus fewer WF cycles are available for matched filtering. The 5-fold increase in the mass ratio errors is also due to the known degeneracy between component mass and (aligned) spins~\cite{2013PhRvD..87b4035B}.
Table~\ref{Tab.BBH_pe} shows the same trends we saw for BNS. In particular, the errors on the chirp mass get larger in the \SQ network as compared to the \BS network, whereas Q is estimated more precisely. However it is still the case that the improvement in the chirp mass estimation is smaller when compared to systematics. We will come back to this point in the next section.
Neither network was able to measure the (aligned) spin magnitudes well, with relative errors being of the order of 70\% on average for both networks.

Even though BBH are not expected to be luminous in the electromagnetic spectrum, making an EM follow-up program of BBH less interesting than for BNS, for completeness we report in Table~\ref{Tab.BBH_pe} the size of the sky error regions.  The average (90,67)\% confidence areas of sky location were (21,11) \sqdeg for the \BS network and (15,7.3) \sqdeg for the \SQ network, corresponding to relative decreases of 29\% and 34\%, respectively with respect to the baseline network.  

Given the similarities between the conclusions drawn about the BNS and BBH events from the comparison of \BS and \SQ, we do not repeat the simulations with the \LS and \LL networks for BBH events. 

\subsubsection{Systematics}\label{SubSubSec.Systematics}

Throughout this paper, we have made claims that the systematic uncertainties in \mc are larger than our measured uncertainties, thus partially limiting the importance of the changes on \mc errors among the detector networks.  While previous studies showed that systematics are indeed the dominating source of error for well-measured BNS chirp masses~\cite{2013PhRvD..88f2001A}, the larger uncertainties in \mc for BBH events are not so clearly negligible compared to waveform systematic errors. 
We have verified if this is the case by re-running the parameter estimation simulations for all BBH events in the \BS network scenario using an IMRPhenomC (IMRc) waveform model~\cite{2010PhRvD..82f4016S} to recover the signal instead of IMRb. 
The idea is that the systematic differences between two different waveform approximants should be somewhat representative of the difference between any of them and the ``real'' gravitational wave signals.

After re-analyzing all BBH signals with IMRc, we estimate the systematic waveform error by comparing the median \mc values for both WF families. The posterior distributions of the chirp mass for a \mc = 11.43 \msun system obtained with the two IMR models is shown in Fig.~\ref{Fig.SystemPost}. For that particular event, the medians are separated by 2.8 standard deviations. 

\bef
\includegraphics[width=0.5\textwidth]{\PlotPath/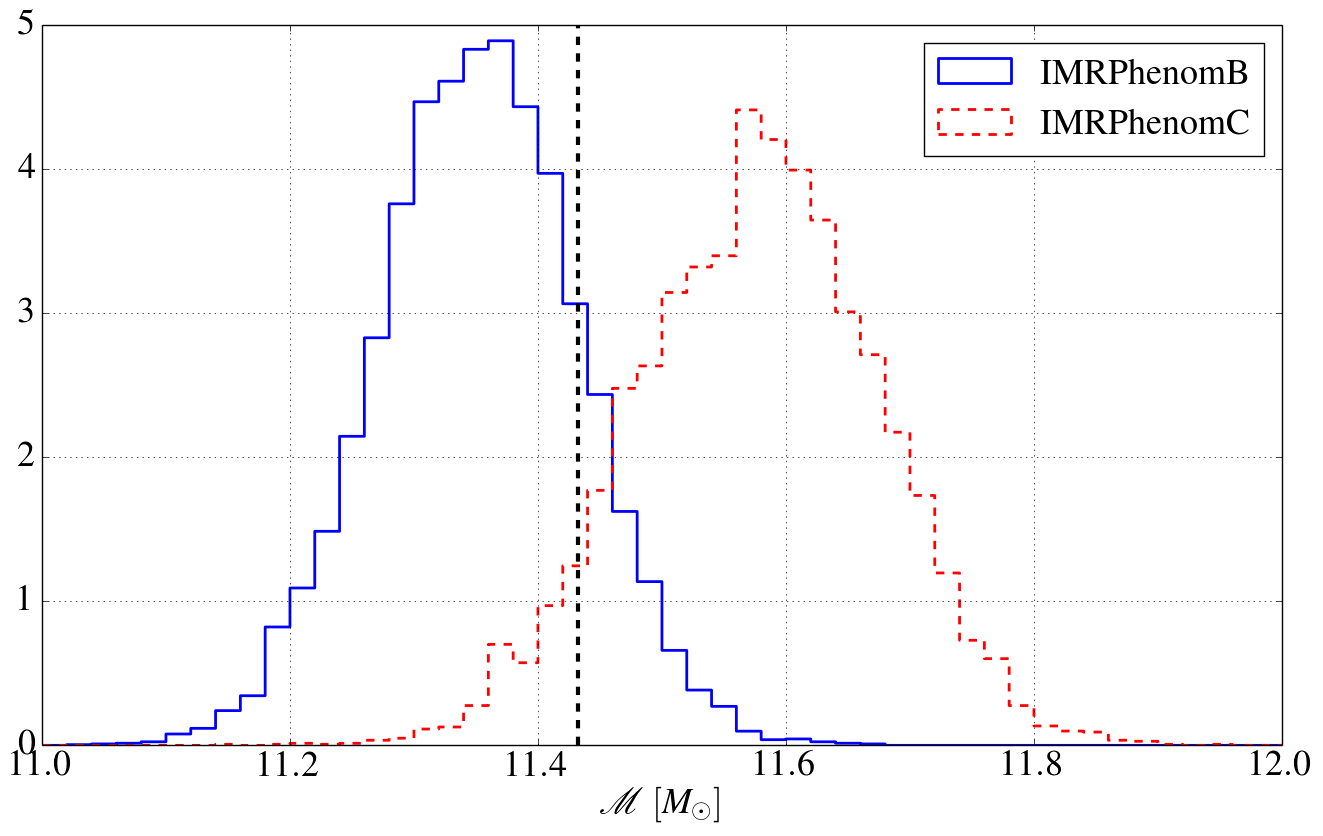}
\caption{The posterior density distributions for \mc obtained with both the IMRb and IMRc waveforms for a BBH event with \mc = 11.43 \msun (analyzed with the \BS network). The absolute separation of the distribution medians is 2.8 times larger than the standard deviation of the IMRb mode.}\label{Fig.SystemPost}
\eef

On average, we found the difference in $\mathcal{M}_{median}$ to be 1.4 standard deviations (using the standard deviation calculated with the IMRb runs). This is an average 1.2\% offset with respect to the injected value. We also found the difference in $Q_{median}$ to be 0.65 standard deviations, which is a 20\% offset with respect to the injected value.  A quick comparison with the relative random errors given in Table~\ref{Tab.BBH_pe} shows how systematics and statistical errors are comparable, and thus the different performances of the \SQ and \BS networks for the estimation of the chirp mass are indeed negligible.

\section{Conclusions}\label{Sec.Conclusions}

In this paper we analyzed the impact of squeezed states of light on the estimation of parameters for gravitational waves emitted by compact binaries and detected by LIGO and Virgo.
We considered a baseline network, where LIGO and Virgo had their design sensitivities\cite{2013arXiv1304.0670L} and compared it with three networks where the LIGO instruments used squeezed light.
These three networks represent plausible implementations of squeezing in GW detectors: a simple frequency-independent squeezer without a filter cavity (\SQ); a squeezer with a lossy filter cavity (\LS) and a squeezer with a lossless filter cavity (\LL).

We showed that for binary neutron star signals the \BS and \SQ networks have essentially the same sensitivity, which means that a simple squeezer will not yield more detections. On the other hand, we found that the sky localization is improved by ~30\% with the \SQ network, which could prove to be a non-negligible improvement for the electromagnetic follow-up of GW events. We also showed that the statistical measurability of neutron star tidal deformability increases by ~30\% with the implementation of frequency-independent squeezing.  Though current waveforms may not be totally reliable at the PN order at which the tidal parameters enter, we expect that this improvement is representative of what can achieved with future, more reliable waveforms. Since frequency-independent squeezing degrades the sensitivity of the instruments in the low-frequency region, these improvements come at the cost of a $\sim$15\% decrease in the measurability of chirp mass; however this loss is effectively negligible compared to limiting systematic waveform errors.

We also showed that implementations of both lossy and lossless frequency-dependent squeezing improve the overall network sensitivity, increasing the average distance of BNS events by 16\% and 28\% respectively. Due to the large boost in LIGO's sensitivity, the SNR of the average event detected by the \LS and \LL networks will mostly be accumulated by the LIGO observatories, degrading the networks' sky localization capabilities. 
We showed how sky localization for events at fixed SNR stays the same for the \LS network and worsens by ~15\% for the \LL network as compared with the \BS network. Finally, for these networks too, it is the case that the measurability of mass parameters is comparable to the \BS network once one takes into account that statistical errors for those parameters are smaller than systematics.

We also analyzed signals emitted by binary black holes, for which we found similar numerical values of the improvements (or degradation) due to squeezing.
BBH events were analyzed using two different waveform families (IMRPhenomB and IMRPhenomC) to get explicit estimates of potential systematics introduced by the WF approximant. We found that on average the medians of the chirp mass estimated with the two WFs are $1.4~\sigma$ away. The effect is smaller for the mass ratio, where the medians are separated by an average of $0.7~\sigma$. 

In summary, implementing quantum squeezing can be expected to increase the scientific payoff of ground based gravitational wave detectors even it its simplest, frequency-independent version, though frequency-dependent squeezing will be required to increase the network's detection rate.

\section{Acknowledgments}

The authors acknowledge the support of the National Science Foundation and the LIGO Laboratory. LIGO was constructed by the California Institute of Technology and Massachusetts Institute of Technology with funding from the National Science Foundation and operates under cooperative agreement PHY-0757058.
The authors would like to acknowledge the LIGO Data Grid clusters, without which the simulations could not have been performed. Specifically, these include the Syracuse University Gravitation and Relativity cluster, which is supported by NSF awards PHY-1040231 and PHY-1104371. Also, we thank the Albert Einstein Institute in Hannover, supported by the Max-Planck-Gesellschaft, for use of the Atlas high-performance computing cluster.
We would also like to thank W.~Del Pozzo, R.~Essick, E.~Katsavounidis, J.~Miller, A.~Weinstein, and the Parameter estimation and CBC groups of the \lvc for useful comments and suggestions.
This is LIGO document number P1400220-v1.

\appendix

\section{\LS and \LL: Same Distance}\label{App.SameD}

In Sec.~\ref{SubSubSec.LossyLossless_SSNR} we analyzed the BNS events considered for the \BS network, changing the distance of each of them so that the network SNR would be the same in all networks.
We found that the sky localization precision does not improve using frequency-dependent squeezing, but rather stays the same (\LS) or worsens (\LL).
As already underlined, this happens because we considered events at a fixed network SNR: since the two LIGO detectors get more sensitive while Virgo (in our simulations) stays the same, the SNR is much more unevenly distributed across the network for the \LS and \LL networks, which negatively affects sky localization.

Another interesting question is what would happen if the \emph{same} events were to be detected with these more sensitive networks, i.e., to compare the effects of squeezing for equal-distance rather than equal-SNR events. 
In this appendix we re-analyze the BNS events keeping their distances (and everything else) to be identical to the \BS configuration.

The main findings are summarized in Table~\ref{Tab.BNS_sameD}. 

\begin{table*}[ht]
\centering
\scriptsize
\caption{The average parameter estimation capabilities of the lossy and lossless networks for BNS events identical to those of the baseline network (i.e., at the same distance).  Note that these results are averaged over all considered BNS events, and that the distribution of these events is not the expected distribution of detected events for the lossy and lossless networks.}\label{Tab.BNS_sameD}
\begin{tabular}[c]{c||c|c|c||c|c||c|c}
PSD & $\Gamma_\mathcal{M}$ (\%) & $\Delta_Q$ (\%) & $\sigma_{t_H}$ (ms) & $\sigma_{t_L}$ (ms) & $\sigma_{t_V}$ (ms) & 90\% Conf (\sqdeg) & 67\% Conf (\sqdeg)\\
\hline\hline
Lossy & $1.1\times10^{-2}$ & 6.1 & 0.16 & 0.19 & 1.5 & 13 & 6.2\\
Lossless & $0.95\times10^{-2}$ & 5.7 & 0.15 & 0.20 & 1.4 & 12 & 6.1\\
\end{tabular}
\end{table*} 

We see that the sky (90\%, 60\%) confidence regions of sky location are (13, 6.2) \sqdeg for the \LS network and (12, 6.1) \sqdeg for the \LL network. These represent a \si30\% improvement with respect to the \BS network. This improvement is due to both the higher SNR and the better high-frequency sensitivity. Furthermore, the fact that the \LS and \LL networks have only slightly smaller timing errors (0.15~ms and 0.16~ms, respectively) than the \SQ network (0.18~ms) despite being significantly better at low-frequency sensitivity, shows again that low frequencies only contribute marginally to time triangulation and thus sky localization. 

We also see from Table~\ref{Tab.BNS_sameD} that the \LS and \LL networks have better mass estimation capabilities than the \BS network for identical events.  The \LS network has the same low-frequency sensitivity as the \BS network and is more sensitive in the most sensitive region. Its average uncertainty in \mc is 21\% lower than the \BS. The \LL network is more sensitive than the \BS network in the whole band, and its uncertainty in \mc is 32\% better than the \BS network. These results indicate that the precision in the measurement of \mc does not only come from the tens of Hertz region, but over a range of frequencies extending into the most sensitive bandwidth of the baseline network. 
Nevertheless, these improvements to the measurement of \mc are again negligible compared to current waveform systematic errors.  Finally, we see that the measured uncertainty in Q is 6.1\% for the \LS network and 5.7\% for the \LL network, corresponding to relative improvements of 20\% and 25\%, respectively, with respect to the \BS network.  We have seen earlier that Q is best estimated at higher frequencies than \mc but at lower frequencies than sky position.  The  results of this appendix confirm such a statement, as the measurability of Q improves as a result of both increases in the high-frequency sensitivity (i.e., in going from the \BS network to the \LS network) and increases in low-frequency sensitivity (i.e., in going from the \LS network to the \LL network). 

\bibliography{./Bibs/pe}

\end{document}